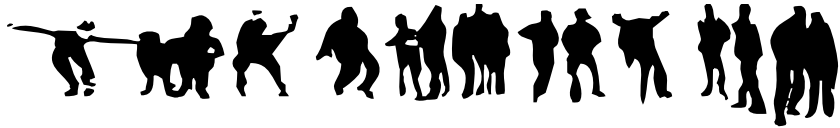

# Etude des déterminants psychologiques de la persistance dans l'usage d'un jeu sérieux : évaluation de l'environnement optimal d'apprentissage avec Mecagenius®.

▶ **Jean HEUTTE** (équipe Trigone-CIREL EA 4353, Lille1 et ESPE Lille Nord de France), **Michel GALAUP** (UMR EFTS MA 122, Toulouse2 Le Mirail), **Catherine LELARDEUX, Pierre LAGARRIGUE, (Institut Clément Ader LGMT EA 814 et CUFR Champollion), Fabien FENOUILLET** (Chart EA 4004, Paris-Ouest Nanterre - la Défense.)

■ **RÉSUMÉ** •L'objectif de cet article est de mettre en évidence que des concepts clés de la motivation sont pertinents pour évaluer l'usage d'un serious game. Les résultats de l'étude exploratoire menée auprès d'étudiants (N=115) impliqués dans l'expérimentation in situ, in vivo de Mecagenius® (jeu dédié aux formations en génie mécanique) permettent aussi confirmer la pertinence de l'usage de l'échelle de mesure du flow en éducation (EduFlow) pour l'évaluation de l'expérience optimale d'apprentissage avec un serious game. Il apparaît par ailleurs qu'EduFlow est bien en relation avec des mesures spécifiques au contexte scolaire comme l'auto-efficacité, l'intérêt, et le climat motivationnel.

■ **MOTS-CLÉS** • Serious game, auto-efficacité, intérêt, climat motivationnel, flow.

■ **ABSTRACT** • *The aim of this paper is to show the relevance of motivational key concepts in evaluating the use of serious game. Thisresearch involves115studentstrainingwith Mecagenius® (serious game in mechanical engineering). The results of theexploratorystudy also confirmthe relevance of theuseof flow in Education scale (EduFlow) to evaluatethe optimallearning experiencewitha serious game. It also appearsthatEduFlowisrelated tospecific actions within the school context such as self-efficacy, motivational climateand interest.*

■ **KEYWORDS** • *Serious game, flow, self-efficacy, interest, motivational climate.*

**Jean HEUTTE, Michel GALAUP, Catherine LELARDEUX,
Pierre LAGARRIGUE, Fabien FENOUILLET**

**Introduction**

En contexte de formation, parmi les objets pédagogiques *ludo-éduquants* (Kellner, 2010), le serious game[1] (Alvarez, 2007), occupe désormais une place particulière. Au delà du simple plaisir de jouer, cette catégorie du jeu vidéo souhaite donner un sens supplémentaire que le seul divertissement (Alvarez et Djaouti, 2010) ; (Lavergne et al., 2010), afin de répondre à des besoins publicitaires, éducatifs, médicaux, environnementaux, etc. Ainsi, l'une des propriétés du serious game serait de susciter l'envie d'apprendre (Alvarez, 2007). Selon Fenouillet et ses collègues. (Fenouillet et al., 2009), cette idée « d'aider » l'envie d'apprendre ne date pas d'hier puisque les jeux vidéo étaient déjà utilisés dans cette optique dès la fin des années 60 (Dorn, 1989). Depuis, cette problématique suscite de nombreux questionnements qui interpellent tout autant les concepteurs de ce type de jeux que les enseignants et chercheurs s'intéressant à l'impact de leurs usages en contexte éducatif.

L'objectif de l'étude exploratoire qui va être présentée dans cet article est de mettre en évidence que des concepts clés de la motivation et de la volition sont pertinents pour évaluer d'une part le déclanchement du comportement motivant l'usage d'un jeu sérieux, d'autre part la persistance de l'usage des jeux sérieux en contexte éducatif. En effet, il nous semble pertinent de considérer la volition comme complément indispensable de la motivation (Fenouillet, 2012), distinguant ainsi les déterminants de la décision d'agir (motivation) de ceux qui éclairent la persistance de l'action (volition). Pour y parvenir nous commencerons dans un premier temps par définir les différents concepts tels que le flow et certains déterminants du climat de classe (auto-efficacité, intérêt et climat motivationnel) que nous avons souhaité retenir pour cette étude. Enfin, nous conclurons en présentant les premiers résultats de l'expérimentation *in situ*, *in vivo* de Mecagenius®, un jeu sérieux dédié aux formations en génie mécanique. Cela nous permettra notamment de mettre en évidence que l'échelle de flow en éducation (EduFlow) élaborée par Heutte, Fenouillet, Boniwell, Martin-Krumm et Csikszentmihalyi (Heutte et al., 2014) est un outil particulièrement bien adapté pour étudier la persistance et l'expérience optimale d'apprentissage avec un jeu sérieux en contexte scolaire/universitaire.

### 1. L'expérience optimale (ou état de flow)

Parmi tous les éléments ayant une influence sur la qualité de l'expérience de jeu des utilisateurs, force est de constater que certains concepts semblent partager des caractéristiques communes : ils englobent pratiquement tous les sphères émotionnelles, sensorielles, cognitives et comportementales. En outre, bien souvent les concepts proposés sont fortement liés à l'idée d'une activité « plaisante » apportant une forme de « satisfaction », en tout cas pour le moins « agréable », les termes revenant le plus régulièrement étant : « divertissement » (*entertainment*), « amusement » (*fun*) et « plaisir » (*enjoyment*). C'est très certainement l'une des raisons pour laquelle la théorie de l'expérience optimale ou l'autotélisme[2] (Csikszentmihalyi, 1990) est de plus en plus souvent présente dans la littérature scientifique et professionnelle concernant la création des jeux vidéo, car provoquer l'état de flow est exactement ce que les concepteurs cherchent à offrir aux joueurs : cela contribue à augmenter à la fois leur plaisir et leur persistance dans le jeu.

Csikszentmihalyi définit le concept de l'expérience optimale qu'il appelle "flow" et qui réfère à l'état subjectif de se sentir bien (Csikszentmihalyi et Patton, 1997). Le flow peut être ressenti dans divers domaines tels que l'art, l'enseignement, le sport... Le flow se manifeste souvent quand il y a perception d'un équilibre entre ses compétences personnelles et la demande de la tâche (*cf.* Figure 1, p. 3) : c'est la perception de cet équilibre « optimal » (entre défit et habilité) qui fait de l'expérience de flow une « expérience optimale ». Dans un jeu vidéo, le flow est, par exemple, ressenti quand l'ensemble des actions à réaliser pour jouer, notamment celles qui réclament une attention soutenue, semblent « couler de source » avec une telle fluidité qu'à aucun moment le jeu ne devra être interrompu par une quelconque inquiétude concernant ce qu'il faut réaliser ou encore concernant les commandes à exécuter pour parvenir.

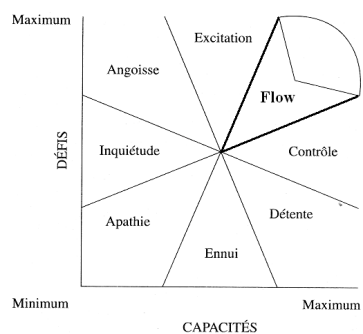

**Figure 1 • Exigences de la tâche et compétences élevées (Csikszentmihalyi, 1990)**

**Jean HEUTTE, Michel GALAUP, Catherine LELARDEUX,
Pierre LAGARRIGUE, Fabien FENOUILLET**

Dans sa définition originale (Csikszentmihalyi, 1990), l'expérience optimale comporte neuf caractéristiques majeures (Tableau 1, p. 4) : (1) équilibre entre défi et habilité, (2) concentration sur la tâche, (3) cible claire, (4) rétroaction claire et précise, (5) absence de distraction, (6) contrôle de l'action, (7) absence de préoccupation à propos du soi – dilatation de l'ego, (8) altération de la perception du temps et (9) bien être.

| | | |
|---|---|---|
| 1 | Équilibre entre défi et habilité | Savoir que l'activité est réalisable, que les compétences sont adéquates, ne ressentir ni angoisse ni ennui, savoir que nous avons toutes les chances de pouvoir terminer l'activité. |
| 2 | Concentration sur la tâche | Être en mesure de se concentrer sur ce que nous faisons. |
| 3 | Cible claire | Grande clarté intérieure, savoir ce qui doit être fait et comment cela pourra se réaliser. La tâche a des objectifs clairs. |
| 4 | Rétroaction claire et précise | La tâche (ou l'environnement) fournit une rétroaction immédiate. |
| 5 | Absence de distraction | Totalement impliqué, concentré, que ce soit en raison d'une grande curiosité, de l'habitude liée à un entraînement important. |
| 6 | Contrôle de l'action | Percevoir que nous pouvons contrôler nos actions. |
| 7 | Absence de préoccupation à propos de soi – dilatation de l'ego[3] (mais paradoxalement, le sens de soi se trouve renforcé) | Sentiment de sérénité, pas de soucis à propos de soi, le sentiment d'une dilation de soi au-delà des frontières habituelle de l'ego - se sentir transcendé au-delà de ce qui nous semblait possible. |
| 8 | Altération de la perception du temps | Le sens de la durée du temps est altéré : les heures semblent être passées en quelques minutes et les minutes peuvent s'étendre à ressembler à des heures. |
| 9 | Expérience autotélique – bien être | Jubilation et extase, sentiment d'être en dehors de la réalité quotidienne, quelque soient les raisons ou les buts de l'action, le flow en est sa propre récompense. L'activité n'est entreprise pour aucune autre raison que le bien-être qu'elle procure. |

**Tableau 1 • Les neuf caractéristiques de l'expérience optimale –** *flow*,
**d'après Csikszentmihalyi (2004), traduction (Heutte, 2011, p. 99)**

Par nature, l'expérience optimale exige une concentration totale de l'attention sur la tâche en cours, ce qui a pour effet d'occulter les aspects déplaisants de la vie, les frustrations ou les préoccupations quotidiennes. Dans le cadre d'un jeu, l'expérience optimale est plus particulièrement ressentie dans les phases qui nécessitent une importante mobilisation de compétences. Par exemple, celles au cours desquelles l'échec ou la réussite tiennent à peu de choses (comme si l'action se déroulait sur le fil d'un rasoir), mais que compte tenu de ses expériences antérieures, le joueur réalise que le défi est très certainement à sa portée. Ainsi, au cours d'une phase de jeu, au fur et à mesure que le joueur s'aperçoit qu'il réussit à s'approcher du but espéré, ce sentiment le portera et le poussera à s'appliquer de plus en plus, en lui procurant un tel bien-être, qu'il souhaitera que cette expérience se prolonge. C'est d'ailleurs pour continuer à ressentir le flow qu'il persistera dans le jeu, y compris parfois pour refaire certaines phases de jeu en se fixant lui-même de nouveaux défis : faire plus vite ou faire mieux, par exemple en économisant les gestes, les actions ou les ressources à sa disposition. Dans la mesure où elle éclaire les déterminants psychologiques de la persistance à agir, l'autotélisme est classée parmi les théories volitionnelles (Fenouillet, 2012 ; Heutte, 2011) : le flow l'un des concepts majeurs pour étudier (ou prédire) la persistance dans le temps d'un comportement humain.

L'expérience optimale entraîne des conséquences très importantes : meilleure performance, créativité, développement des capacités, estime de soi et réduction du stress (Csikszentmihalyi, 2006). Un ensemble d'études (pour revue, voir (Delle Fave et al., 2011) apportent des résultats concourants et montrent l'importance d'autres concepts dans l'expérience du flow. Par exemple, Asakawa (Asakawa, 2004) met en évidence des liens positifs entre la motivation et le flow, ainsi que des liens négatifs entre le flow et l'anxiété ou le désengagement. Pour notre part, nous considérons qu'un environnement optimal d'apprentissage est un environnement d'apprentissage qui soutient le flow (l'expérience optimale) en contexte éducatif (Heutte, sous presse).

### 1.1. Le flow dans l'éducation

Progressivement, un nombre grandissant de recherches s'intéressent à l'impact du flow en contexte éducatif, par exemple pour étudier, la motivation des élèves de collèges (Rathunde et Csikszentmihalyi, 2005), dans des lycées (Peterson et Miller, 2004) ou dans des universités (Shernoff et al., 2003) ; (Heutte, 2011), tout particulièrement dans des contextes de formation à distance (Caron et al., 2014) ; (Heutte, Caron et al., 2014) ; (Heutte, Kaplan et al. 2014) . Turner et Meyer (2004) ont notamment étudié l'impact du soutien et de l'étayage des étudiants par les enseignants sur le flow.

### 1.2. Time Flies When You're Having Fun

De longue date, la plupart des tentatives d'explications du comportement individuel des utilisateurs de technologies de l'information et de la communication (TIC) ont tendance à se concentrer essentiellement sur les croyances de maîtrise instrumentale pour comprendre leurs intentions

**Jean HEUTTE, Michel GALAUP, Catherine LELARDEUX,
Pierre LAGARRIGUE, Fabien FENOUILLET**

d'usage des TIC. Cependant, (Choi et al.,2007) ainsi que (Pearce et al., 2005) font état du grand intérêt et du caractère prometteur des recherches concernant le flow dans les environnements numériques. En effet, le flow est une variable très régulièrement évoquée pour comprendre les expériences positives avec les ordinateurs (pour revue, voir (Heutte, 2011), et plus récemment, pour ce qui concerne l'usage d'internet (Chen, 2000). Cette théorie a notamment été utilisée afin de mieux appréhender l'immersion pendant les activités d'exploration (Ghani, 1995), de communication (Trevino et Webster, 1992), et d'apprentissage (Ghani, 1995).

Pour leur part, Agarwal et Karahanna (Agarwal et Karahanna, 2000) proposent le concept d'absorption cognitive (AC) qu'ils définissent comme un profond état d'engagement à travers cinq dimensions : (1) dissociation temporelle (perte de la notion du temps) ; (2) immersion (concentration totale) dans une tâche ; (3) intensité du plaisir ; (4) sentiment de contrôle de l'interaction ; (5) curiosité sensorielle et cognitive. Ces épisodes d'attention totale qui « absorbent » entièrement les ressources cognitives au point que plus rien d'autre n'importe sont des expériences optimales, des états de flow (Agarwal et Karahanna, 2000). L'AC et ses cinq dimensions sont des antécédents significatifs de la perception d'utilité et celle de la convivialité (cf. Figure 2, p. 6) : deux dimensions empruntées au Technology Acceptance Model (TAM) (Davis, 1989) ; (Venkatesh et Davis, 2000). L'AC est donc un état spécifique qui résulte à la fois de facteurs individuels et situationnels. Elle renforce l'intention d'utiliser les technologies numériques, elle serait de plus particulièrement bénéfique au sentiment de réalisation d'un individu dans le cadre de son travail et, par conséquent, influencerait sa motivation.

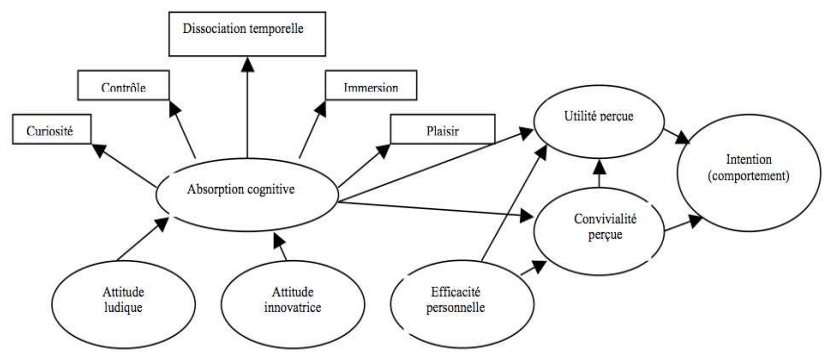

**Figure 2 • Le modèle de l'absorption cognitive d'après Agarwal et Karahanna (2000, p. 683)**

### 1.3. L'absorption cognitive : quand plus rien ne peut perturber.

En état de flow, les personnes sont tellement concentrés dans l'activité que plus rien d'autre ne peut les perturber. Au-delà du plaisir lié à l'activité et de la persistance liée à l'intérêt intrinsèque pour l'activité, l'immersion totale dans l'activité semble être un aspect central de l'expérience de flow.

Pour notre part (Heutte, sous presse), nous suggérons de reprendre l'idée-force du concept d'AC pour l'étendre à toute activité centré sur la volonté de comprendre avec, comme sans, l'usage des technologies numériques. Nous estimons que l'AC « correspond à un profond état d'engagement lié à un épisode d'attention totale (expérience optimale) qui « absorbe » (qui focalise) entièrement les ressources cognitives au point que plus rien d'autre n'importe que de comprendre, ce qui a notamment pour conséquence immédiate que pratiquement plus rien ne peut effectivement perturber la concentration exclusivement centrée sur la compréhension » (Heutte, 2011, p. 105).

### 1.4. La mesure du flow

L'une des difficultés de l'étude du flow réside en grande partie dans son identification ainsi que sa mesure. Ainsi, depuis plus d'une trentaine d'années, divers outils ayant des formes variées ont été créés afin d'étudier les éléments de nature instable et les phénomènes subjectifs liés au flow : aussi paradoxal que cela puisse paraître, malgré un large consensus concernant la définition du flow, les désaccords entre chercheurs concernant sa mesure sont l'objet de très nombreuses controverses (Moneta, 2012). C'est une des raisons principales de l'émergence de nombreuses recherches afin de tenter de mesurer le flow à l'aide de questionnaires (pour revue, voir (Heutte 2011)). Deux des échelles les plus utilisées sont la version 2 du Flow State Scale (FSS-2) et du Dispositional Flow Scale(DFS-2), élaborées par Jackson et Eklund (Jackson et Eklund, 2002), initialementen vue de mesurer le flow en contexte sportif, sur la base des 9 dimensions conceptuelles originelles du flow. Ces outilsont depuis été utilisés dans d'autres contextes. Cependant, Procci et ses collègues (Procci et al. 2012) mettent en évidence d'importantes limites à leur utilisation en contexte d'usage des jeux vidéos. Enfin, avant les travaux d'Heutte, Fenouillet, Boniwell, Martin-Krumm et Csikszentmihalyi (Heutte, Fenouillet et al., 2014) pour l'élaboration de l'échelle EduFlow, il n'existait pas d'échelle courte spécifique pour l'étude des dimensions et des effets du flow en contexte éducatif.

**Jean HEUTTE, Michel GALAUP, Catherine LELARDEUX,
Pierre LAGARRIGUE, Fabien FENOUILLET**

## 2. Le climat de classe

Nous référant au courant récent de la psychologie positive (Seligman etCsikszentmihalyi, 2000) qui est « l'étude scientifique des conditions et des processus qui contribuent à l'épanouissement ou au fonctionnement optimal des individus, des groupes et des institutions » (Gable et Haidt, 2005), p. 103, il nous semble nécessaire d'intégrer la dimension psychosociale du contexte d'apprentissage ou climat de classe, afin d'étudier les éventuelles différences liées à l'usage d'unjeu sérieux.

### 2.1. L'auto-efficacité

La théorie de l'auto-efficacité (Bandura,1997) entre dans le cadre plus large de la théorie sociale cognitive (Bandura, 1986). Selon cette théorie, l'auto-efficacité ou sentiment d'efficacité personnelle constitue la croyance que possède un individu en sa capacité de produire ou non une tâche : « l'auto-efficacité (personnelle comme collective) renvoie aux jugements que les personnes font à propos de leur(s) capacité(s), personnelle(s) comme collective, à organiser et réaliser des ensembles d'actions requis pour atteindre des types de performances attendus mais aussi aux croyances à propos de leurs capacités à mobiliser la motivation, les ressources cognitives et les comportements nécessaires pour exercer un contrôle sur les événements de la vie » (Heutte, 2011), p. 95. Plus grand est le sentiment d'auto-efficacité, plus élevés sont les objectifs que s'impose la personne et l'engagement dans leur poursuite. Les croyances d'auto-efficacité des élèves sont positivement associées à l'acquisition directe des apprentissages et à la réussite scolaire, mais également *via* la préférence pour des tâches présentant un certain niveau de nouveauté.

### 2.2. L'intérêt

L'étude des liens entre l'intérêt et la motivation fait partie des traditions en psychologie (e.g., (Dewey, 1913) ; (Thorndike, 1935)), elle se trouve renouvelée et actualisée par de nouvelles approches théoriques, comme le modèle développement par phases de l'intérêt (Hidi et Renninger, 2006). Hidi et Renninger suggèrent que non seulement les intérêts se développent mais, qu'en plus, ils ne sont pas forcément stables. Pour mieux en étudier les phases de développement, elles proposent de distinguer *l'intérêt situationnel* (par exemple, l'intérêt lié à la situation pédagogique provoquée par l'usage d'un jeu en classe) de *l'intérêt individuel* (par exemple, l'intérêt pour une discipline académique (géométrie euclidienne, littérature contemporaine, génétique, histoire de l'antiquité, génie méca-

nique…) en tant que telle, en dehors même d'un contexte scolaire particulier). Ainsi, selon ces chercheuses, le développement se réaliserait dans cet ordre : (1) le déclanchement (l'activation) de l'intérêt situationnel, (2) le maintien de l'intérêt situationnel, (3) l'émergence de l'intérêt individuel, (4) l'intérêt individuel développé. L'intérêt situationnel serait à la base de l'intérêt individuel.

### 2.3. Le climat motivationnel scolaire

Pour Sarrazin et ses collègues (Sarrazin et al., 2006), le climat motivationnel instauré par l'enseignant a un impact important sur l'implication des élèves en classe. De nombreux auteurs (pour revue, voir (Heutte, 2011)) soulignent l'importance du soutien des besoins psychologiques de base (Deci et Ryan, 2002) ; (Deci et Ryan, 2008), le climat de classe serait ainsi tout particulièrement influencé par le sentiment d'appartenance sociale des élèves (avec leurs collègues, comme avec leurs enseignants). Le *modèle sociocognitif des apprentissages scolaires* (Leroy et *al.,* 2013) souligne le rôle médiateur fondamental des perceptions de l'environnement social. Autrement dit, « ce n'est pas la réalité objective qui compte le plus, mais bien la manière dont l'individu – l'élève en l'occurrence – la construit. » (Leroy et *al.,* 2013), p. 12, notamment sa perception du soutien des enseignants. Ainsi, le climat motivationnel scolaire est un construit résultant principalement des perceptions (nécessairement subjectives) que les élèves ont des actions objectivement réalisées par les enseignants.

Ces travaux récents sont d'une très grande complémentarité avec ceux concernant l'importance de l'auto-efficacité (sentiment d'efficacité personnelle et collective), de l'intérêt et du flow dans l'expérience optimale d'apprentissage (Bandura, 2003) ; (Deci & Ryan, 2002) ; (Deci et Ryan, 2008) ; (Heutte, 2011), élevant le niveau de défis acceptés par les étudiants (Meyer et Turner, 2006), ainsi que leur persistance dans les apprentissages (Shernoff et Csikszentmihalyi, 2009). C'est la raison pour laquelle il nous a semblé opportun d'intégrer quelques variables du climat de classe dans cette étude, en vue d'en étudier les éventuels effets d'interaction dans un contexte pédagogique lié à l'usage d'un jeu sérieux.

**Jean HEUTTE, Michel GALAUP, Catherine LELARDEUX,
Pierre LAGARRIGUE, Fabien FENOUILLET**

### 3. Contexte de l'étude exploratoire

### 3.1. Expérimentation Mecagenius®*in situ* et *in vivo*

Fruit d'une collaboration entre le groupement d'intérêt scientifique (GIS) Serious Game Research Network sous la responsabilité scientifique du centre universitaire Champollion et l'industriel KTM Advance, Mecagenius® est un jeu sérieux en ligne (*cf.* Figure 3).

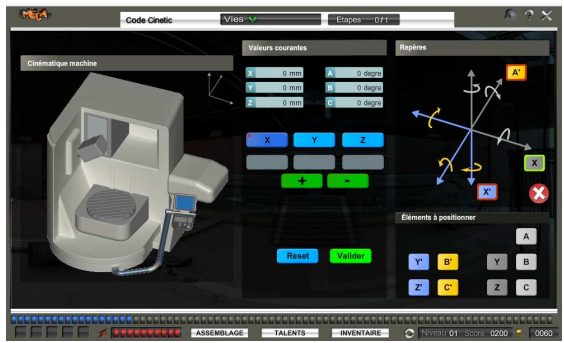

**Figure 3 • Mecagenius® : un jeu sérieux en génie mécanique
pour découvrir, apprendre, fabriquer**(http://mecagenius.univ-jfc.fr/ )

Mecagenius® intègre plus de 200 activités pédagogiques au sein d'une aventure spatiale futuriste et reproduit l'univers d'une industrie de génie mécanique allant de la découverte de l'atelier de génie mécanique à l'optimisation de production sur machine outil à commande numérique (Galaup, 2013) ; (Galaup et al., 2010) ; (Lelardeux et al , 2011). L'expérimentation *in situ* et *in vivo* en Île-de-France est la première phase de test grandeur nature. Les régions Île-de-France et Midi Pyrénées, les pôles Aerospace Valley et ASTech participent au financement de l'utilisation de Mecagenius® dans une quinzaine d'établissements d'enseignement secondaires et supérieurs. Ces expérimentations ont débuté en Février 2013 et se poursuivent en 2014. La première étude exploratoire (dont il est rendu compte dans les pages suivantes) était principalement destinée à tester le protocole en vue de la prochaine phase de l'expérimentation de Mecagenius® qui a lieu dans les académies de Créteil, de Lille et de Toulouse au cours de l'année scolaire/universitaire 2013/2014. En Île-de-France, l'expérimentation devait concerner des élèves de lycées professionnels et des étudiants en licence et en master. En raison de quelques soucis techniques et de difficultés de planification des activités avec Mecagenius® dans certains établissements, le corpus de données concerne majoritairement des étudiants de L3 et de M1 en formation à l'École Nationale Supérieure d'Arts et Métiers. Comme on peut s'y attendre pour ce type de formation, les étudiants sont majoritairement de sexe masculin (91,1%), et d'un âge moyen de 21,7 ans. Les phases de jeux organisées par l'enseignant se sont déroulées au cours de séances de travaux pratiques, en petits groupes (24 étudiants maximum).

### 3.2. Évaluation de l'expérimentation

L'évaluation de cette expérimentation souhaite croiser différents points de vue (*cf.* évaluation « kaléidoscopique » (Lelardeux et al., 2013), notamment en didactique (Galaup, 2013) ; (Galaup, Lelardeux et al., 2012), ainsi qu'en psychologie des apprentissages, *via* un test de validité écologique du *modèle heuristique du collectif individuellement motivé* (MHCIM) (Heutte, 2011) en contexte d'usage de jeu sérieux (Heutte, 2012).

Le travail de recherche en didactique (Galaup, 2013) a consisté à s'intéresser aux phénomènes de transposition didactique (Schubauer-Leoni et Leutenegger, 2005) ; (Amade-Escot, 2007) liés à la conception, puis à l'usage d'un jeu sérieux dans les classes. Une des spécificités de ce travail a été de s'appuyer sur le cadre conceptuel de la théorie de l'action conjointe en didactique (Sensevy et Mercier, 2007) ; (Amade-Escot et Leutenegger, 2013) ; (Leutenegger, 2005) permettant de fournir à la fois des outils pour la conception de Mecagenius® et pour l'analyse de ces usages, dans la perspective d'alimenter la réflexion et les recherches sur les jeux sérieux. Nous avons proposé d'identifier et de comprendre les processus mis en œuvre par les acteurs, professeur et élève(s) lorsqu'ils utilisent l'artefact Mecagenius® dans le cadre d'un enseignement et d'un apprentissage des savoirs du génie mécanique. Ce travail a participé à la construction d'une connaissance scientifique des pratiques telles qu'elles existent dans leur diversité et a permis d'éclairer les processus d'apprentissage mis en œuvre à partir d'une analyse *in situ* de l'utilisation de Mecagenius® en contexte d'enseignement (Galaup, 2013) ; (Galaup, Amade-Escot et al., 2012) ; (Lelardeux et al., 2010). L'évaluation de l'usage au regard de la théorie de l'action conjointe en didactique fait d'ailleurs l'objet d'un article spécifique (Galaup et Amade-Escot, 2013). Cependant, nous avons considéré que l'étude des dynamiques d'apprentissage des élèves devait être complétée par l'étude de l'intérêt, du bien-être et de la motivation des joueurs à persister à vouloir apprendre avec Mecagenius®. En effet, l'analyse des stratégies de jeu des élèves doit être regardée selon ces deux axes car le bien-être psychologique au travers du jeu est un catalyseur de la dynamique d'apprentissage. Les pages qui suivent seront donc principalement centrées sur les déterminants psychologiques de l'expérience optimale avec Mecagenius® : elles souhaitent rendre compte de l'étude exploratoire destinée à mettre en évidence que certains outils de mesure des concepts clés de la motivation sont pertinents pour évaluer l'usage d'un jeu sérieux.

### 3.3. Protocole d'évaluation des déterminants psychologiques de l'apprentissage avec Mecagenius®

Avant même la présentation de Mecagenius®, tous les étudiants avaient à répondre à un questionnaire *ante*, basé en grande partie sur les échelles de mesure composant le MHCIM (Heutte, 2011) : sentiment d'appartenance sociale (Richer et Vallerand, 1998), sentiment d'efficacité personnelle (adaptation française de (Schwarzer et Jerusalem, 1995)), EduFlow (Heutte et al., 2014). Dans la mesure où il s'agissait d'une première démarche exploratoire pour un test de validité écologique du MHCIM dans ce contexte (jeu sérieux et génie mécanique), certaines échelles supplémentaires ont été agrégées au questionnaire en vue de d'établir d'autres comparaisons (validité concourante) : L'échelle de l'Intérêt Situationnel et Individuel pour les Serious Games (IS2G)(Chainon et al. 2014), la version française de Flow State Scale-2 (FSS2)(Fournier et *al.*, 2007), la version adaptée et traduite du Learning Climate Questionnaire (LCQ) (William & Deci, 1996) afin de pouvoir évaluer le style de l'enseignant perçu par les élèves (Leroy et al., 2013).

Chaque étudiant jouait seul sur son poste de travail (Galaup et Amade-Escot, 2013). Après chaque séance de jeu, ils étaient invités à répondre à un court questionnaire en ligne. Les 14 étudiants inscrits en BAC PRO technicien d'usinage et les 43 étudiants inscrits en filière Génie Énergétique (M1) on participé à une séance de jeu (ils ont donc été invités à répondre à un total de 2 questionnaires), les 46 étudiants inscrits en filière Génie Industriel (L3) ont pu participer à 2 séances (ils ont donc été invités à répondre à un total de 3 questionnaires).

### 3.4. Instruments

Tous les outils est constitués de façon identique : une liste d'items pour lesquels les répondants devaient indiquer s'ils étaient « tout à fait d'accord » ou « pas du tout d'accord » selon une échelle de Likert en sept pas.

### 3.4.1. Mesure du flow

Dans la mesure où pour étudier le flow il est préférable d'utiliser une échelle courte et compte tenu de certaines réserves concernant la qualité de la mesure du flow sur la base de ces 9 dimensions conceptuelles originelles en contexte d'usage de jeu vidéo (Procci et al., 2012), dans cette étude exploratoire, nous avons souhaité dans un premier temps comparer la version française de FSS2 (Fournier et al., 2007) avec EduFlow (Heutte et al., 2014). Comme EduFlow présente l'intérêt d'être à la fois compacte (12 items) et multidimensionnelle (en 4 dimensions), c'est cette échelle qui a été privilégiée par la suite pour les mesures répétées du flow après chaque séance de jeu.

– FSS2 (Fournier et al., 2007) composée de 32 items répartis en 9 sous-échelles de 4 items :

- FSS2D1-Équilibre entre défi et habilité (eg. « *Mes capacités sont à la hauteur du défi élevé de la situation* »)
- FSS2D2- Fusion de l'action et de la conscience (eg. « *Je fais les bonnes actions sans essayer d'y penser* »)
- FSS2D3- Cible claire (eg. « *Je sais clairement ce que je veux faire* »)
- FSS2D4-Rétroaction claire et précise (eg « *J'ai une vision très claire de ce je veux faire* »)
- FSS2D5-Concentration (eg. « *Je suis entièrement concentré sur ce que je fais* »)
- FSS2D6- Contrôle de l'action (eg. « *J'ai l'impression de contrôler ce que je fais* »)
- FSS2D7 - Absence de préoccupation à propos du soi (eg. « *Je ne suis pas préoccupé par mon apparence* »)
- FSS2D8- Altération de la perception du temps (eg. « *Je perds ma notion habituelle du temps* »)
- FSS2D9- Bien-être (eg. « *J'aime vraiment cette expérience* »)

– EduFlow (Heutte *et al.*, 2014) composée de 12 items répartis en 4 sous échelles de 3 items :

- EduFlowD1- Absorption cognitive (eg. « *A chaque étape, je sais ce que je dois faire* ».)
- EduFlowD2- Altération de la perception du temps (eg. « *Je ne vois pas le temps passer* »)
- EduFlowD3 - Absence de préoccupation à propos du soi (eg. « *Je ne suis pas inquiet de ce que les autres peuvent penser de moi* »)
- EduFlowD4- Bien-être scolaire (eg. « *J'ai le sentiment de vivre un moment enthousiasmant* »)

**Jean HEUTTE, Michel GALAUP, Catherine LELARDEUX,
Pierre LAGARRIGUE, Fabien FENOUILLET**

### 3.4.2. Mesure du climat de classe

Nous avons ensuite souhaité comparer les différentes composantes du flow avec d'autres variables qui constituent des indicateurs du climat de classe, pour ce faire nous avons retenu l'auto-efficacité, l'intérêt et le climat motivationnel scolaire.

#### 3.4.2.1. Mesure de l'auto-efficacité

Une phrase introductive permettant de contextualiser la passation pour deux outils de mesure de l'auto-efficacité : « *Généralement, dans le cadre des activités liées à mon travail scolaire (en cours comme en dehors des cours)...* » :

– Echelle du sentiment d'efficacité personnel (ESEP, adaptation française de (Schwarzer et Jerusalem, 1995)) composée de 10 items (eg. « *Grâce à mes compétences, je sais gérer des situations inattendues* »)

– Echelle du sentiment d'efficacité collective (ESEC, Piguet, 2008) composée de 10 items (eg. « *Quoiqu'il arrive, nous savons généralement faire face* »)

#### 3.4.2.2. Mesure de l'intérêt

Nous inspirant du modèle de *développement par phase de l'intérêt* (Hidi et Renninger, 2006), nous avons souhaité distinguer *l'intérêt situationnel* (dans le cas présent, l'intérêt lié à la situation provoquée par Mecagenius®) de *l'intérêt individuel* (dans le cas présent, l'intérêt pour le génie mécanique en tant que tel).

Pour mesurer ces différentes dimensions de l'intérêt, nous avons utilisé l'échelle de l'Intérêt Situationnel et Individuel pour les Serious Games (IS2G)(Chainon et al., 2014) qui est constitué de 3 sous échelles :

– Interet_ind - Intérêt individuel (6 items) (eg. « *J'aime le génie mécanique* »)

– Interet_SA - Intérêt situationnel activé (4 items) (eg. « *Mecagenius® est si intéressant qu'il capte facilement l'attention* »)

– Interet_SM - Intérêt situationnel maintenu (4 items) (eg. « *Je trouve que tout ce que m'a appris Mecagenius® peut être utile* »)

#### 3.4.2.3. Mesure du climat motivationnel scolaire

Pour appréhender le climat motivationnel scolaire, nous avons utilisé deux outils de mesure :

– l'échelle du sentiment d'appartenance sociale (ESAS, Richer et Vallerand, 1998) que nous avons utilisé pour mesurer les perceptions liées aux étudiants et celles liées aux enseignants se décompose en deux sous-échelles de 4 items :

- ESAS_etud_A – Sentiment d'acceptation avec les étudiants (eg. « *Dans mes relations avec mes collègues, je me sens estimé(e)* »)

- ESAS_etud_I – Sentiment d'intimité avec les étudiants (eg. « *Dans mes relations avec mes collègues, je sens un(e) ami(e) pour eux* »)
- ESAS_ens_A – Sentiment d'acceptation avec les enseignants (eg. « *Dans mes relations avec mes enseignants, je me sens écouté(e)* »)
- ESAS_ens_I – Sentiment d'intimité avec les enseignants (eg. « *Dans mes relations avec mes enseignants, je me sens attaché(e) à eux* »)

– le questionnaire du climat d'apprentissage (version adaptée et traduite du learning climate questionnaire (LCQFr ; Leroy et *al.*, 2013) permettant de mesurer le style motivationnel de l'enseignant perçu par les élèves, *via* 7 items. (eg. « *Je sens que mes professeurs m'acceptent comme je suis* »)

## 4. Résultats

Tout d'abord, nous constatons qu'après les différentes phases de jeux les étudiants ont progressivement été de moins en moins nombreux à jouer, mais aussi à renseigner les questionnaires (participation libre).

Avant les séances de jeux (T0) :
- effectif théorique : N=115
- questionnaires complets exploitables : N= 73 (63,47%)

1$^{ère}$ séance de jeux (T1) :
- effectif théorique : N=115
- questionnaires complets exploitables : N= 67 (58,26%)

2$^e$ séance de jeux (T2) :
- effectif théorique : N=43
- questionnaires complets exploitables : N= 26 (60,46%)

Nous pouvons trouver trois niveaux d'explications à ce constat. Premièrement, en tant que telle l'expérimentation a démarré plus tardivement que prévu (forte proximité avec la fin d'année scolaire/universitaire). De ce fait, il était parfois difficile pour les enseignants de programmer effectivement toutes les séances initialement planifiées. Ensuite, force est de constater que la multitude des protocoles correspondant à des points de vue scientifiquement complémentaires (*cf.* didactique, psychologie et sociologie) ont généré un peu de confusion chez les enseignants d'où parfois des difficultés de compréhension des méthodologies et des exigences spécifiques concernant les différents pro-tocoles mis en œuvre. Toutes les consignes n'ont pas toujours pu être rappelées à temps auprès des étudiants : de ce fait, certains étudiants ont oublié de répondre au questionnaire en fin de séance de jeu. Enfin, un souci technique (panne de l'un des serveurs d'enquête) en cours d'expérimentation a malheureusement perturbé la collecte de données pendant 4 jours, entre T1 et T2.

### 4.1. Qualité des outils de mesure

|  | Valid N | Mean | Std.Dev. | Skewness | Kurtosis | Alpha |
|---|---|---|---|---|---|---|
| Interet_ind_T0 | 108 | 3,79 | 1,17 | .16 | -.12 | .87 |
| EduFlowD1_T0 | 107 | 3,92 | 1,25 | .06 | -.26 | .86 |
| EduFlowD2_T0 | 107 | 3,84 | 1,42 | .09 | -.48 | .81 |
| EduFlowD3_T0 | 107 | 5,04 | 1,57 | -.69 | -.08 | .92 |
| EduFlowD4_T0 | 106 | 3,53 | 1,28 | .13 | -.05 | .83 |
| FSS2D1_T0 | 102 | 4,03 | 1,21 | -.26 | -.19 | .92 |
| FSS2D2_T0 | 100 | 3,63 | 1,25 | -.29 | -.76 | .92 |
| FSS2D3_T0 | 102 | 4,16 | 1,27 | -.39 | -.39 | .91 |
| FSS2D4_T0 | 102 | 4,12 | 1,28 | -.40 | -.28 | .91 |
| FSS2D5_T0 | 101 | 4,49 | 1,26 | -.75 | .63 | .93 |
| FSS2D6_T0 | 102 | 3,95 | 1,22 | -.40 | -.11 | .93 |
| FSS2D7_T0 | 102 | 4,88 | 1,56 | -.64 | -.13 | .89 |
| FSS2D8_T0 | 102 | 4,22 | 1,41 | -.33 | -.33 | .88 |
| FSS2D9_T0 | 102 | 4,04 | 1,21 | -.42 | .00 | .90 |
| ESEP_T0 | 104 | 5,00 | .91 | -.26 | -.54 | .92 |
| ESEC_T0 | 103 | 5,31 | .89 | -.51 | .16 | .95 |
| ESAS_etud_A_T0 | 103 | 4,84 | .88 | -.18 | -.07 | .86 |
| ESAS_etud_I_T0 | 103 | 4,67 | 1,16 | -.49 | .06 | .92 |
| ESAS_enseig_A_T0 | 102 | 4,16 | 1,02 | -.20 | -.09 | .88 |
| ESAS_enseig_I_T0 | 102 | 2,86 | 1,23 | .44 | -.56 | .93 |
| LCQFr_T0 | 105 | 3,27 | .62 | -.17 | -.24 | .81 |
| EduFlowD1_T1 | 100 | 4,58 | .93 | .04 | .27 | .77 |
| EduFlowD2_T1 | 100 | 4,79 | 1,38 | -.38 | -.23 | .86 |
| EduFlowD3_T1 | 100 | 5,27 | 1,61 | -.74 | -.29 | .92 |
| EduFlowD4_T1 | 100 | 4,10 | 1,32 | -.15 | -.08 | .90 |
| Interet_ind_T1 | 99 | 3,72 | 1,10 | .32 | .01 | .88 |
| Interet_SA_T1 | 99 | 4,66 | 1,20 | -.30 | -.07 | .86 |
| Interet_SM_T1 | 99 | 4,66 | 1,21 | -.25 | -.15 | .86 |

**Tableau 1 • Statistiques descriptives et consistances internes des échelles et dimensions utilisées**

Nous apprécions (*cf*. Tableau 1) la consistance interne des échelles à l'aide du coefficient alpha de Chronbach (1951). Celui-ci permet de tester la cohérence de chaque item avec l'ensemble des autres énoncés de l'échelle à laquelle il appartient. Tous les coefficients alpha sont supérieurs à .77, ce qui indique une fiabilité tout à fait acceptable de nos outils (Nunnally, 1978). Nous avons également apprécié la symétrie (Skewness), ainsi que l'aplatissement (Kurtosis) de chaque distribution. Tous nos coef-ficients sont compris entre -.75 et .63 ce qui nous permet d'estimer que ces distributions ne sont pas trop éloignées des critères d'une normalité univariée. Pour Kline (2010) aucun de ces indicateurs ne dépasse les seuils de non normalité univariée qui peuvent être qualifiés de sévères.

### 4.2. EduFlow : un outil de mesure adapté au contexte universitaire

#### 4.2.1. Comparaisons entre EduFlow et FSS2

Avant d'entreprendre les analyses liées à l'évaluation de la première phase exploratoire de Mecagenius®*in situ* et *in vivo* en tant que telle, il nous a semblé nécessaire de nous assurer que l'échelle de flow en éducation (EduFlow) pouvait être avantageusement utilisée à la place de la version française de Flow State Scale-2 (FSS2). Nous avons aussi souhaité comparer ces échelles de mesure du flow avec celles de l'intérêt individuel académique, en 6 items (Chainon et al. 2014) et du sentiment d'efficacité personnelle, ESEP en 10 items (adaptation française de (Schwarzer et Jerusalem, 1995)), car ces variables sont souvent associées aux déterminants motivationnels du flow (Bassi et al., 2007) ; (Delle Fave et al., 2011) ; (Heutte, 2011).

Dans un premier temps, nous constatons que l'analyse factorielle en composantes principales avec rotation varimax de l'échelle FSS2 (*cf.* Tableau 2) met en évidence 5 facteurs (et non 9 comme attendu) avec des valeurs propres supérieures à 1 (qui expliquent 76,9% de la variance cumulée).

Les deux premiers facteurs concentrent les 6 premières dimensions de FSS2 postulées par Jackson et Eklund (Jackson et Eklund, 2002), avec cependant près d'un tiers d'items ambigus :

- 1er facteur (*cf.* Tableau 2),principalement : FSS2-D1, "Équilibre entre défi et habilité", FSS2-D2 "fusion de l'action et de la conscience", FSS2-D3 "cible claire", FSS2-D4 "rétroaction" et FSS2-D6 "contrôle de l'action",
- 2e facteur, principalement : FSS2-D5 "concentration"

Le 3e et le 4$^e$ facteurs (*cf.* Tableau 2) mettent assez clairement en évidence deux dimensions de FSS2 postulées par Jackson et Eklund (Jackson et Eklund, 2002):

- 3e facteur : FSS2-D8 "Altération de la perception du temps"
- 4e facteur : FSS2-D7 "Absence de préoccupation à propos du soi"

Enfin, malgré de très nombreux items ambigus, le 5e facteurs (*cf.* Tableau 2) met principalement en évidence FSS2-D9 "bien-être / expérience autotélique".

| N=80 | \multicolumn{5}{c|}{Composantes} |
|---|---|---|---|---|---|
| | 1 | 2 | 3 | 4 | 5 |
| FSS2-D1a | .78 | | | | |
| FSS2-D1b | .80 | | | | |
| FSS2-D1c | .73 | .47 | | | |
| FSS2-D1d | .76 | | | | |
| FSS2-D2a | .79 | | | | |
| FSS2-D2b | .79 | | | | |
| FSS2-D2c | .75 | | | | |
| FSS2-D2d | .76 | | | | |
| FSS2-D3a | .83 | | | | |
| FSS2-D3b | .81 | | | | |
| FSS2-D3c | .49 | | .53 | | |
| FSS2-D3d | .70 | .47 | | | |
| FSS2-D4a | .84 | | | | |
| FSS2-D4b | .79 | | | | |
| FSS2-D4c | .69 | .45 | | | |
| FSS2-D4d | .58 | | | | |
| FSS2-D5a | | .75 | | | |
| FSS2-D5b | .45 | .67 | | | |
| FSS2-D5c | | .83 | | | |
| FSS2-D5d | | .80 | | | |
| FSS2-D6a | .80 | | | | |
| FSS2-D6b | .61 | .60 | | | |
| FSS2-D6c | .73 | .43 | | | |
| FSS2-D6d | .64 | .55 | | | |
| FSS2-D7a | | | | .90 | |
| FSS2-D7b | | | | .86 | |
| FSS2-D7c | | | | .75 | |
| FSS2-D7d | | | | .90 | |
| FSS2-D8a | | | .87 | | |
| FSS2-D8b | | | .86 | | |
| FSS2-D8c | | | .42 | | .65 |
| FSS2-D8d | | | .87 | | |
| FSS2-D9a | .53 | | | | .52 |
| FSS2-D9b | .48 | .44 | | | |
| FSS2-D9c | .48 | .44 | | | .56 |
| FSS2-D9d | .59 | | | | |
| % variance | 53.72 | 8.62 | 6.21 | 5.15 | 3.21 |
| Valeur propre | 19.34 | 3.10 | 2.24 | 1.86 | 1.16 |

Méthode d'extraction : maximum de vraisemblance.
Les items avec de faibles contributions factorielles (valeurs inférieures à .40)
ont été éliminés (Hair et al., 1995).

**Tableau 2 • Analyse en composantes principales de l'échelle FSS2**

*Sticef, Numéro Spécial Évaluation dans les Jeux Sérieux* – **Recueil 2014**Les résultats (*cf.* Tableau 3) mettent en évidence une corrélation significative et forte[4] ($r = .75$*[5]) entre le score obtenu en réalisant la moyenne de l'ensemble des 12 items d'EduFlow et des 36 items de la version française de FSS2.fr.

| *N*=73 | FSS2 global | EduFlow global | EduFlow D1 | EduFlow D2 | EduFlow D3 | EduFlow D4 |
|---|---|---|---|---|---|---|
| FSS2 global | 1.00 | .75* | .69* | .60* | .46* | .60* |
| FSS2-D1 | .92* | .70* | .72* | .55* | .40* | .58* |
| FSS2-D2 | .87* | .68* | .70* | .56* | .31* | .59* |
| FSS2-D3 | .94* | .69* | .66* | .52* | .45* | .55* |
| FSS2-D4 | .93* | .66* | .69* | .50* | .38* | .53* |
| FSS2-D5 | .80* | .50* | .47* | .48* | .22* | .44* |
| FSS2-D6 | .88* | .61* | .64* | .49* | .35* | .47* |
| FSS2-D7 | .53* | .45* | .29* | .24* | .62* | .20* |
| FSS2-D8 | .73* | .62* | .45* | .62* | .35* | .48* |
| FSS2-D9 | .88* | .69* | .59* | .62* | .33* | .69* |
| interet_ind | .61* | .69* | .66* | .62* | .26* | .72* |
| ESEP | .31* | .40* | .40* | .35* | .15* | .41* |

**Tableau 3 • Corrélations entre les dimensions des échelles EduFlow et FSS2**
(* = corrélation significative *p*<.05)

La dimension 1 d'EduFlow, l'absorption cognitive (EduFlowD1) et certaines dimensions de FSS2 (*cf.*Tableau 3) entretiennent des liens significatifs et importants pour toutes les dimensions qui lui correspondent : FSS2D1-Équilibre entre défi et habilité ($r = .72$*), FSS2D2-Fusion de l'action et de la conscience ($r = .70$*), FSS2D3-Cible claire ($r = .66$*), FSS2D4-Rétroaction claire et précise ($r = .69$*), FSS2D5-Consentration ($r = .47$*), FSS2D6- Contrôle de l'action ($r = .64$*).

Les dimensions qui mesurent l'altération de la perception du temps, respectivement EduFlowD2 et FSS2D8 (*cf.*Tableau 3) sont significativement liées avec une intensité importante ($r = .62$*). Les dimensions qui mesurent l'absence de préoccupation à propos du soi, respectivement EduFlowD3 et FSS2D7 entretiennent des liens significatifs et importants ($r = .62$*). Les dimensions qui mesurent le bien-être procuré par l'activité en tant que telle, respectivement EduFlowD4 et FSS2D9 (*cf.*Tableau 3) entretiennent elles aussi des liens significatifs et importants ($r = .69$*). Enfin, nous pouvons noter qu'EduFlow est plus fortement liée à l'intérêt individuel académique des élèves que FSS2 ($r = .69$* *vsr* = .61*), ainsi qu'à leur sentiment d'efficacité personnelle dans les études en génie mécanique ($r = .40$* *vsr* = .31*).

**Jean HEUTTE, Michel GALAUP, Catherine LELARDEUX,
Pierre LAGARRIGUE, Fabien FENOUILLET**

#### 4.2.2. Le flow en contexte d'usage de Mecagenius®

Comme annoncé précédemment, nous avons procédé à deux mesures du flow :

– T0 avant la première séance de TD avec Mecagenius®, la phrase introductive permettant de contextualiser la passation de l'échelle EduFlow était : « *Quand je dois réaliser une activité en génie mécanique (cinématique, conception, métrologie, productique, usinage…)…* »

– T1 juste après la séance de TD avec Mecagenius®, la phrase introductive permettant de contextualiser la passation de l'échelle EduFlow était : « *Quand j'utilise Mecagenius®…* »

Ainsi, nous pouvons considérer que la première mesure (T0) correspond à la mesure du flow dans une activité en génie mécanique « habituelle » (sans jeu) et que la deuxième mesure (T1) correspond à la mesure du flow dans une activité en génie mécanique avec Mecagenius®. Les résultats (*cf.* Tableau 4) mettent en évidence des corrélations significatives et globalement moyennes entre trois dimensions du flow avec ou sans l'usage de Mecagenius® (.33* < r < .36*). Concernant EduFlowD4, le bien-être, cette relation est significative et importante (r = .47*). La relation moyenne des liens entre la plupart des dimensions du flow nous permet d'en déduire que bien que comportant de nettes similitudes en terme de construction (en 4 dimensions), le flow ressenti par les étudiants pendant une séance avec Mecagenius® n'est pas exactement le même que celui qu'ils ressentent pendant une séance habituelle de génie mécanique (puisque globalement, quand elles sont significatives, les corrélations ne sont pas très fortes).

| $N$=67 | EduFlow D1_T1 | EduFlow D2_T1 | EduFlow D3_T1 | EduFlow D4_T1 |
|---|---|---|---|---|
| EduFlow_D1_T0 | .33* | .01 | .05 | .18 |
| EduFlow_D2_T0 | .20 | .34* | .16 | .26* |
| EduFlow_D3_T0 | .18 | .07 | .36* | .15 |
| EduFlow_D4_T0 | .37* | .23 | .17 | .47* |

**Tableau 4 • Corrélations entre les dimensions d'EduFlow
avec (T1) ou sans jeu sérieux (T0)**
(* = corrélation significative *p*<.05)

La comparaison des moyennes (*cf.* Tableau 5) d'EduFlow (T0=3.94 *vs* T1=4.71) met en évidence une augmentation globale (+19.58%.) significative ($t(72) = -5.62$, $p<.001$) du flow en contexte d'une activité en génie mécanique avec Mecagenius®. Des analyses plus détaillées mettent en évidence que c'est pour l'altération de la perception du temps (EduFlowD2) que la progression (T0=3.72 *vs* T1=4.95, soit +33.17%) est la plus importante (*t*(72) = - 6.71 , *p*<.001).

| N = 73 | Flow | Moy. | S.D. |
|---|---|---|---|
| sans Jeu sérieux (T0) | EduFlow_D1 | 3,82 | 1,17 |
| | EduFlow_D2 | 3,72 | 1,34 |
| | EduFlow_D3 | 4,86 | 1,63 |
| | EduFlow_D4 | 3,37 | 1,17 |
| | **EduFlow global** | **3,94** | **1,01** |
| avec Jeu sérieux (T1) | EduFlow_D1 | 4,53 | 0,90 |
| | EduFlow_D2 | 4,95 | 1,31 |
| | EduFlow_D3 | 5,22 | 1,72 |
| | EduFlow_D4 | 4,15 | 1,22 |
| | **EduFlow global** | **4,71** | **0,97** |

**Tableau 5 • Statistiques descriptives des dimensions d'EduFlow avec (T1) ou sans jeu sérieux (T0)**

Ensuite, il apparaît que c'est pour le bien-être (EduFlowD4) que la progression (T0=3.37 *vs* T1=4.15, soit +23.24%) est significative (*t*(72) = - 5.41 , *p*<.001). Pour l'absorption cognitive (EduFlowD1), la progression (T0=3.82 *vs* T1=4.53, soit +18.61%) est elle aussi significative (*t*(72) = - 5.01 , *p*<.001). Cependant, cette différence (T0=4.86 *vs* T1=5.22) n'est pas significative (*t*(72) = - 1.66 , ns) pour ce qui concerne l'absence de préoccupation à propos du soi (EduFlowD3).

#### 4.2.3. Le flow et les variables du climat motivationnel scolaire

##### 4.2.3.1. Le flow et l'auto-efficacité personnelle/collective

Globalement (*cf*. Tableau 6), le flow est significativement moyennement lié au sentiment d'efficacité personnelle (*r* = .40\*) et, dans une moindre mesure au sentiment d'efficacité collective (*r* = .23\*). Une analyse plus détaillée confirme que pour toutes les dimensions, l'effet, quand il est significatif, est systématiquement plus élevé entre le flow et le sentiment d'efficacité personnelle (*vs* sentiment d'efficacité collective). Dans un ordre décroissant, l'effet est le plus remarquable pour EduFlowD4, le bien-être (*r* = .41\* *vsr* = .32\*), EduFlowD1, l'absorption cognitive (*r* = .40\* *vsr* = .22\*) et EduFlowD2, l'altération de la perception du temps (*r* = .35\* *vsr* = .23\*).

**Jean HEUTTE, Michel GALAUP, Catherine LELARDEUX,
Pierre LAGARRIGUE, Fabien FENOUILLET**

| $N$=67 | EduFlow global | EduFlow D1 | EduFlow D2 | EduFlow D3 | EduFlow D4 |
|---|---|---|---|---|---|
| interet_ind | .69* | .66* | .62* | .26* | .72* |
| ESEP | .40* | .40* | .35* | .15 | .41* |
| ESEC | .23* | .22* | .23* | .01 | .32* |
| ESAS_etud_A | .20 | .10 | .19 | .16 | .19 |
| ESAS_etud_I | .21* | .09 | .18 | .17 | .23* |
| ESAS_ens_A | .37* | .38* | .25* | .21* | .32* |
| ESAS_ens_I | .39* | .35* | .36* | .14 | .40* |
| LCQFr | .30* | .39* | .25* | .11 | .31* |

**Tableau 6• Corrélations entre les dimensions d'EduFlow
et différentes échelles liées au climat motivationnel**
(* = corrélation significative *p*<.05)

*4.2.3.2. Le flow et la qualité du climat de classe*

Les résultats (*cf.* Tableau 6) mettent en évidence de nombreux liens significatifs, d'une intensité moyenne entre le sentiment d'appartenance sociale et le flow. Globalement le flow est davantage lié au sentiment d'appartenance sociale avec les enseignants (acceptation $r$ = .37* et intimité $r$ = .39*), qu'avec les autres étudiants (acceptation $r$ = n.s. et intimité $r$ = .21*). En affinant l'analyse, il apparaît que le sentiment d'appartenance sociale vis à vis des enseignants est plus fortement lié avec EduFlowD1, l'absorption cognitive (acceptation $r$ = .38* et intimité $r$ = .35*) et EduFlowD4, le bien-être (acceptation $r$ = .32* et intimité $r$ = .40*), qu'avec EduFlowD2, l'altération de la perception du temps (acceptation $r$ = .25* et intimité $r$ = .36*). EduFlowD3, l'absence de préoccupation à propos du soi n'entretient qu'un lien significatif faible avec le sentiment d'acceptation vis à vis des enseignants ($r$ = .21*). La qualité du climat d'apprentissage entretient des liens significatifs faibles avec presque toutes les dimensions du flow (à l'exception de EduFlowD3, l'absence de préoccupation à propos du soi). C'est avec EduFlowD1, l'absorption cognitive que le lien est le plus élevé ($r$ = .39*).

*4.2.3.3. Le flow et l'intérêt*

Comme nous pouvions nous y attendre, les résultats mettent en évidence que le flow en contexte de jeu n'est pas systématiquement lié à l'intérêt individuel pour le génie mécanique (*cf.* Tableaux 7 et 8). C'est pour ce qui concerne l'intérêt situationnel activé et l'altération de la perception du temps (EduFlowD2) que les liens sont les plus remarquablement élevés (T1 : $r$ =.68* (*cf.* Tableau 7) et T2 : $r$ =.84* (*cf.* Tableau 8)).

| N=67 | interet ind_T0 | Interet ind_T1 | Interet SA_T1 | Interet SM_T1 |
|---|---|---|---|---|
| EduFlowD1_T1 | .33* | .52* | .45* | .44* |
| EduFlowD2_T1 | .06* | .35* | .68* | .40* |
| EduFlowD3_T1 | .11* | .32* | .16* | .05* |
| EduFlowD4_T1 | .28* | .56* | .60* | .44* |

**Tableau 7 • Corrélations entre les dimensions d'EduFlow, les dimensions de l'intérêt situationnel et individuel avec (T1) ou sans jeu sérieux (T0)**
(* = corrélation significative *p*<.05)

| N=23 | Interet ind_T2 | Interet SA_T2 | Interet SM_T2 |
|---|---|---|---|
| EduFlowD1_T2 | .32* | .48* | .43* |
| EduFlowD2_T2 | .47* | .84* | .52* |
| EduFlowD3_T2 | .19* | .18* | .15* |
| EduFlowD4_T2 | .64* | .67* | .66* |

**Tableau 8 • Corrélations entre les dimensions EduFlow et les dimensions de l'intérêt situationnel et individuel avec jeu sérieux, 2e séance (T2)**
(* = corrélation significative *p*<.05)

| N=67 | interet ind_T0 | Interet ind_T1 | Interet SA_T1 | Interet SM_T1 |
|---|---|---|---|---|
| interet_ind_T0 | 1,00 | .62* | .03* | .05* |
| Interet_ind_T1 | .62* | 1,00 | .42* | .40* |
| Interet_SA_T1 | .03* | .42* | 1,00 | .77* |
| Interet_SM_T1 | .05* | .40* | .77* | 1,00 |

**Tableau 9 • Corrélations entre les différentes les dimensions de l'intérêt situationnel et individuel sans T0 et avec T1**
(* = corrélation significative *p*<.05)

Nous constatons d'une part (*cf.* Tableau 9) qu'à la première séance de jeu (T1) les liens entre l'intérêt individuel (ind T1) et l'intérêt situationnel activé (SA_T1) ou l'intérêt situationnel maintenu (SM_T1) sont significatifs et moyens (respectivement *r* = .42* et *r* =.40*) et que d'autre part (*cf.* Tableau 10) à la deuxième séance de jeu (T2), les liens entre l'intérêt individuel (ind T1) et l'intérêt situationnel activé (SA_T1) ou l'intérêt situationnel maintenu (SM_T1) sont significatifs et plus importants (respectivement *r* = .62* et *r* =.54*), alors que toujours au cours de la deuxième séance de jeu (T2) les liens entre l'intérêt individuel (ind T2) et l'intérêt situationnel activé (SA_T2) ou l'intérêt situationnel maintenu (SM_T2) sont significatifs et encore plus élevés (respectivement *r* = .63* et *r* =.79*).

**Jean HEUTTE, Michel GALAUP, Catherine LELARDEUX,
Pierre LAGARRIGUE, Fabien FENOUILLET**

Cela est d'autant plus remarquable pour ce qui concerne les liens entre l'intérêt situationnel activé (SA) et l'intérêt situationnel maintenu (SM) au cours des différentes phases de jeu (en T1 : r=.40* (*cf.* Tableau 9) et en T2 : r=.79* (*cf.* Tableau 10)).

| N=23 | Interet ind_T1 | Interet SA_T1 | Interet SM_T1 | Interet ind_T2 | Interet SA_T2 | Interet SM_T2 |
|---|---|---|---|---|---|---|
| Interet_ind_T1 | 1,00 | .62* | .54* | .67* | .40* | .63* |
| Interet_SA_T1 | .62* | 1,00 | .66* | .38* | .66* | .52* |
| Interet_SM_T1 | .54* | .66* | 1,00 | .21* | .41* | .47* |
| Interet_ind_T2 | .67* | .38* | .21* | 1,00 | .63* | .79* |
| Interet_SA_T2 | .40* | .66* | .41* | .63* | 1,00 | .68* |
| Interet_SM_T2 | .63* | .52* | .47* | .79* | .68* | 1,00 |

**Tableau 10 • Corrélations entre les différentes les dimensions de l'intérêt situationnel et individuel en fonction des séances de jeux (T1 *vs* T2)**
(* = corrélation significative *p*<.05)

Constatant cette évolution de la répartition de la variance, donc pour ainsi dire l'évolution d'un meilleur pouvoir explicatif des liens entre les différentes formes d'intérêt, nous nous estimons en mesure de formuler l'hypothèse (à confirmer) que Mecagenius® renforce l'intérêt individuel pour le génie mécanique, puisqu'au cours des différentes séances de jeux, l'intérêt individuel (pour le génie mécanique) est lié de façon de plus en plus importante à l'intérêt situationnel maintenu (lié au fait de jouer avec Mecagenius®).

## 5. Discussion

### 5.1. Limites de cette étude

Le faible nombre de répondants réduit la portée de certains résultats de cette étude, cela peut expliquer le fait de ne pas retrouver les 9 dimensions postulées par Jackson et Eklund (Jackson et Eklund, 2002) dans l'échelle FSS2. Cependant, nos résultats vont bien dans le sens de ceux de Procci et ses collègues (Procci et al., 2012) concernant la variabilité des perceptions du flow suivant les contextes, ainsi quela remise en cause de l'incontournabilité des outils de mesure du flow élaborés sur la base de ces 9 dimensions conceptuelles originales. Pour s'en assurer, il conviendrait de mettre en place d'autres études impliquant un nombre plus important de sujets pour étendre la validité écologique et limiter les biais d'échantillonnages.

### 5.2. EduFlow un outil idéal pour la mesure du flow en classe

Nos résultats mettent en évidence que l'échelle de mesure du flow en éducation (EduFlow) et Flow State Scale-2 (FSS2) mesurent des construits présentant globalement d'importantes similitudes. Certaines dimensions comme l'absence de préoccupation à propos du soi (EduFlowD3/FSS2D7), l'altération de la perception du temps (EduFlow-D2/FSS2-D8) ou encore le bien-être (EduFlowD4/FSS2D9) sont de remarquablement équivalentes. EduFlow apparaît donc comme une échelle qui capte globalement le même construit que FSS2, mais avec l'avantage d'être trois fois plus courte (12 items *vs* 36 items). Dans la mesure où 6 des 9 dimensions postulées de FSS2 ne semblent pas être réellement correctement perçues par les apprenants, la conceptualisation très compacte de l'absorption cognitive (EduFlowD1) fait d'EduFlow un outil idéal pour la mesure du flow en contexte éducatif, avec comme sans, l'usage de technologies numériques (Heutte, sous presse).

Deux derniers éléments plaident en faveur de l'usage de l'échelle de flow en éducation : en effet, EduFlow est globalement plus fortement liée à l'intérêt individuel académique des élèves que FSS2, ainsi qu'à leur sentiment d'efficacité personnelle dans les études. De ce fait, EduFlow semble mieux adaptée pour un usage en contexte éducatif, car davantage liée à des variables qui sont généralement considérées comme de bons prédicteurs de la réussite scolaire (Bandura, 1986) ; (Bandura, 2003 ; (Deci et Ryan, 2002) ; (Deci et Ryan, 2008) ; (Fenouillet, 2012) ; (Heutte, 2011); (Schiefele et Csikszentmihalyi, 1994).

Sans pour autant remettre en cause les qualités psychométriques de FSS2, qui reste valide dans de très nombreux contextes, EduFlow apparaît comme une alternative originale et novatrice. A la fois multidimensionnelle et courte, cette échelle de mesure permet de mieux comprendre les mécanismes liés aux effets d'interactions avec les différentes dimensions du flow. Ainsi, l'échelle de flow en éducation (EduFlow) satisfait pleinement aux besoins de l'évaluation dans les jeux sérieux. De part sa facilité de passation et du pouvoir prédictif du flow sur la persistance, EduFlow sera très adaptée pour des études longitudinales, notamment en vue d'optimiser les dispositifs pédagogiques, *just in time*, en cours de conception.

**Jean HEUTTE, Michel GALAUP, Catherine LELARDEUX,
Pierre LAGARRIGUE, Fabien FENOUILLET**

### 5.3. Le flow en contexte d'usage d'un jeu sérieux

Nos résultats mettent en évidence que tout en présentant des liens entre leurs structures (autour de 4 dimensions : absorption cognitive, altération de la perception du temps, absence de préoccupation à propos du soi et bien-être), le flow ressenti dans une situation d'enseignement ordinaire est différent de celui ressenti en contexte d'usage d'un jeu sérieux. Ceci confirme que la perception du flow est variable et contextuelle (Procci et al., 2012). La comparaison des moyennes d'EduFlow met en évidence une progression globale significative du flow dans le contexte d'une activité de formation avec un jeu sérieux.

Comme l'on pouvait s'y attendre (*c.f.* « *Time Flies When You're Having Fun* » (Agarwal et Karahanna, 2000)), c'est au niveau de l'altération de la perception du temps (EduFlowD2) que la progression est la plus importante. Pour le bien-être (EduFlow-D4) comme pour l'absorption cognitive (EduFlowD1), cette progression est significative et remarquable.

A la lumière de ces différents éléments, nous pouvons en conclure que l'introduction d'un jeu sérieux améliore l'expérience optimale d'apprentissage des apprenants. Comme nous pouvions nous y attendre, l'altération de la perception du temps est liée à l'intérêt situationnel provoqué par l'introduction du jeu sérieux au cours de la formation. Nous formulons l'hypothèse (à confirmer dans de prochaines études) que cela contribue à augmenter leur de temps de travail (donc leur persistance), à renforcer leur concentration dans le travail (*via* l'absorption cognitive), tout en leur procurant du bien-être ce qui peut être prédicteur d'un réengagement dans l'activité, afin de percevoir à nouveau l'ensemble des éléments plaisants de cette expérience optimale.

### 5.4. Le flow et le climat motivationnel

#### 5.4.1. Le flow et l'auto-efficacité en contexte éducatif

Nos résultats mettent en évidence des liens significatifs entre le flow et l'auto-efficacité en contexte éducatif : cela est tout à fait conforme aux attentes (Bassi et al., 2007) ; (Fenouillet et al., 2014) ; (Heutte, 2011). Dans le détail nous pouvons noter que l'effet est remarquable pour EduFlowD4, le bien-être, EduFlowD1, l'absorption cognitive et EduFlowD2, l'altération de la perception du temps.

Le fait dans cette première phase de déploiement que Mecagenius® soit utilisé en situation de jeu individuel (*vs* jeu multi-joueurs en réseau) peut être une explication aux constats d'une systématique plus faible intensité pour ce qui concerne les liens entre le flow et le sentiment d'efficacité collective. Cela peut très certainement aussi expliquer l'absence de lien significatif entre l'auto-efficacité et l'absence de préoccupation à propos du soi (EduFlowD3).

Quoi qu'il en soit, compte tenu du pouvoir prédictif de l'auto-efficacité sur la réussite académique (Bandura, 2003), l'étude du flow en contexte éducatif (étayée par l'usage d'EduFlow) semble porteuse de pistes de recherche très prometteuses, notamment afin d'éclairer les contributions des différentes dimensions du flow dans la persistance à vouloir comprendre (Heutte, 2011) ; (Heutte et al., 2014) et en lien avec l'étude des dynamiques d'apprentissage des élèves (Galaup, 2013).

**5.4.2. Flow et intérêt**

Conformément aux attentes (Hidi et Renninger, 2006), nos résultats mettent en évidence que le flow en contexte de jeu n'est pas systématiquement lié à l'intérêt individuel pour la discipline. Comme nous pouvions nous y attendre, les liens sont remarquablement élevés et en progression constante entre l'altération de la perception du temps (EduFlowD2) et l'intérêt situationnel activé, ainsi qu'entre le bien-être (EduFlowD4) et l'intérêt situationnel activé. Cela met clairement en évidence que l'introduction du jeu sérieux déclenche l'intérêt situationnel. Cela est bien conforme aux attentes (Hidi et Renninger, 2006), c'était d'ailleurs tout à fait souhaité par les concepteurs du jeu.

Au cours des différentes phases de l'expérimentation, nous pouvons constater que le lien entre l'intérêt individuel avant et après la première séance de jeu, puis entre la première et la deuxième séance de jeu a légèrement progressé. D'autre part, nous constatons aussi une remarquable progression concernant les liens entre l'intérêt individuel et l'intérêt situationnel maintenu.

Ces faisceaux de concordance concernant l'évolution de la répartition de la variance des résultats observés nous poussent à tenter l'interprétation d'un meilleur pouvoir explicatif des liens entre les différentes formes d'intérêt, par la formulation de l'hypothèse (à confirmer) que le jeu sérieux renforce l'intérêt individuel pour la discipline académique, puisqu'au cours des différentes séances de jeu, l'intérêt individuel est lié de façon de plus en plus importante à l'intérêt situationnel maintenu. Bien entendu, la faiblesse de l'effectif en deuxième séance de jeu, ne permet pas de conclure d'une façon définitive, mais cela semble pour le moins indiquer une tendance qui ouvre sur des perspectives de recherche particulièrement intéressantes.

### 5.4.3. Le flow et la qualité du climat de classe

Nos travaux mettent en évidence les liens entre la qualité du climat d'apprentissage, le sentiment d'appartenance sociale avec les enseignants et l'expérience optimale d'apprentissage. Le lien entre l'absorption cognitive (EduFlowD1) et la qualité du climat d'apprentissage est remarquable. Cela n'est pas surprenant quand on sait que la mauvaise qualité du climat de classe est très souvent évoquée pour expliquer les difficultés des étudiants à pouvoir se concentrer correctement sur leurs taches d'apprentissage.

Du point de vue des étudiants, l'absorption cognitive (EduFlowD1) et le bien-être (EduFlowD4) sont liés au sentiment d'appartenance sociale avec les enseignants, ce qui est conforme aux attentes (Heutte, 2011). Tous ces éléments permettent de rappeler le rôle central de l'enseignant dans la mise en place d'un climat propice aux apprentissages. On peut d'ailleurs relever que l'absorption cognitive (EduFlowD1) est plus fortement liée au fait de se sentir accepté par les enseignants, que de se sentir en proximité avec eux. Pour sa part, le bien-être (EduFlowD4) est plus fortement lié au fait de se sentir en proximité avec les enseignants que de se sentir accepté par eux. Si l'on veut bien considérer d'une part l'importance du sentiment d'acceptation (notamment l'intime conviction que l'enseignant accepte l'erreur de l'étudiant car elle fait partie de l'apprentissage) sur les mécanismes cognitifs et, d'autre part, une dimension plus affective du bien-être psychologique, alors ces deux résultats semblent tout à fait cohérents.

## 6. Conclusion

Cette coopération (didactique - psychologie des apprentissages) avait pour objectif de construire des outils méthodologiques permettant de prendre en compte tout le potentiel du jeu sérieuxainsi que la diversité des utilisations possibles. Le résultat principal de cette étude exploratoire est le constat que les concepts clés de la motivation sont pertinents pour évaluer l'usage d'un jeu sérieux en classe. Globalement, les résultats mettre d'ailleurs en évidence des liens significatifs et parfois élevés entre l'usage du jeu, le flow et le climat de classe (l'auto-efficacité, l'intérêt et le climat motivationnel scolaire). Ainsi, nous pouvons conclure qu'en tant que telle, dans la mesure où elle renforce le flow des apprenants, l'introduction de Mecagenius® dans la formation contribue à la conception d'un environnement optimal d'apprentissage.

Enfin, il apparaît qu'EduFlow (Heutte et al., 2014) est une alternative originale et novatrice, tout à fait adaptée pour l'étude du flow en contexte éducatif. A la fois multidimensionnelle et courte, EduFlow permet d'améliorer les connaissances scientifiques concernant la compréhension de mécanismes liées aux déterminants et aux effets de certaines dimensions du flow en contexte éducatif. De part sa facilité de passation et du pouvoir prédictif du flow sur la persistance, EduFlow sera très adaptée pour des études longitudinales, notamment en vue d'optimiser les dispositifs pédagogiques, *just in time*, sans devoir nécessairement attendre les résultats des évaluations académiques finales pour les améliorer.

Sur un plan pratique/pragmatique, EduFlow peut très utilement contribuer à l'élaboration de tableaux de bord du pilotage de la qualité des dispositifs pédagogiques, notamment dans le cadre de démarches d'amélioration continue (Heutte, Déro et al., 2014). Voilà qui devrait pouvoir utilement contribuer à la conception pédagogique et didactique d'un environnement optimal d'apprentissage adapté aux dispositions de chaque type d'apprenant dans des contextes variés, ainsi qu'à conforter l'étayage scientifique de la pédagogie universitaire (De Ketele, 2010), plus particulièrement dans des contextes les dispositifs pédagogiques instrumentés en réseau : massive online open course (MOOC), communautés épistémiques en ligne, jeux sérieux…

### 7. BIBLIOGRAPHIE

**Jean HEUTTE, Michel GALAUP, Catherine LELARDEUX,
Pierre LAGARRIGUE, Fabien FENOUILLET**

**Jean HEUTTE, Michel GALAUP, Catherine LELARDEUX,
Pierre LAGARRIGUE, Fabien FENOUILLET**

*« Les environnements personnels d'apprentissage. Entre description et modélisation : quelles approches, quels modèles ? »*

**Jean HEUTTE, Michel GALAUP, Catherine LELARDEUX,
Pierre LAGARRIGUE, Fabien FENOUILLET**

---

[1] Selon la définition d'Alvarez un serious game est une « application informatique, dont l'intention initiale est de combiner, avec cohérence, à la fois des aspects sérieux (Serious) tels, de manière non exhaustive et non exclusive, l'enseignement, l'apprentissage, la communication, ou encore l'information, avec des ressorts ludiques issus du jeu vidéo (Game)»(2007, p. 51).

[2] « Autotélique » est un mot composé de deux racines grecques : *autos* (soi-même) et *telos* (but). Une activité est autotélique lorsqu'elle est entreprise sans autre but qu'elle-même. Une grande partie la poésie française à partir de Baudelaire et surtout de Rimbaud revêt diverses formes d'autotélisme. Le poème devient un objet possédant un intérêt en soi hors de toute nécessité de référentialité extérieure : sa finalité est d'être un processus de création poétique en tant que tel. Dans le domaine de la psychologie, l'autotélisme est une modélisation théorique qui souhaite éclairer les comportements lies à la persistance dans une activité qui n'est entreprise sans autre but qu'elle-même, juste pour prolonger le sentiment de bien-être qu'elle procure.

[3] « Paradoxalement, donc, l'ego est dilaté par un type d'action où l'on s'oublie soi-même » (Csikszentmihalyi, 2006, p. 112) dans la traduction française par Claude-Christine Farny de l'ouvrage original « Creativity », paru en 1996.

[4] Nous interprétons nos résultats en référence à Corroyer et Rouanet (1994) :
- autour de .10 "effet faible",
- à partir de .24 "effet moyen" ,
- à partir de .45 "effet important".

[5] Seuils de significativité : * = *p*< .05